\documentclass{ws-procs9x6}
\usepackage{graphicx}
\usepackage{amsmath,amssymb}

\def\nn{\nonumber}

\def\b0{b_0}

\def\cal{\mathcal}

\begin{document}

\title{Threshold Resummation Effects in the polarized 
Drell-Yan process at GSI and J-PARC\footnote{
\uppercase{T}alk presented by \uppercase{H}.\ \uppercase{Y}okoya at the
``\uppercase{XIV} \uppercase{I}nternational \uppercase{W}orkshop
on \uppercase{D}eep \uppercase{I}nelastic \uppercase{S}cattering
(\uppercase{DIS}2006)'', \uppercase{A}pril 20-24, 2006,
\uppercase{T}sukuba, \uppercase{J}apan.}}

\author{Hiroshi Yokoya} 
\address{Department of Physics, Niigata University,\\
Niigata 950-2181, Japan\\
E-mail: yokoya@nt.sc.niigata-u.ac.jp}

\author{Werner Vogelsang}
\address{Physics Department, Brookhaven National Laboratory,\\
Upton, New York 11973, U.S.A.\\
E-mail: vogelsan@quark.phy.bnl.gov}

\maketitle

\abstracts{
We present studies of QCD corrections to dilepton production
in transversely polarized $pp$ and $\bar{p}p$ scattering. 
In particular we briefly discuss the effects of NNLL threshold 
resummation on the rapidity distribution of the lepton pair.}

\begin{picture}(5,2)(-245,-410)
\put(0,-95){BNL-NT-06/21}
\end{picture}

\section{Introduction}

Recently, new experiments in polarized hadron collisions have been
proposed at the GSI\cite{gsi} ($\bar{p}p$) and at J-PARC\cite{jparc} ($pp$). 
One of the main purposes of these experiments is the measurement of
transverse-spin asymmetries in the Drell-Yan process, in order
to get information on the transversely polarized parton distribution
functions (PDFs)
of the nucleon. The proposed experiments would be at relatively modest
collision energies, e.g.\ $\sqrt{S}=14.5$~GeV at GSI-PAX and $\sqrt{S}=
10$~GeV at J-PARC. At these energies, perturbative-QCD (pQCD) corrections
as well as power-suppressed contributions may be large and require
careful theoretical study.

In this brief note, we report on recent studies of pQCD corrections 
to the invariant-mass and rapidity distributions of Drell-Yan 
pairs\cite{shimizu}. In particular, we consider the
all-order resummation of large ``threshold'' logarithms\cite{thre}.
Our aim is to see the behavior of QCD higher-order corrections
in this kinematic regime, and to investigate the self-consistency 
of the pQCD framework. For further details, including a discussion
of possible non-perturbative effects to the cross section, 
see\cite{shimizu}.

\section{Mass Distributions}
The invariant-mass distribution of Drell-Yan lepton pairs
can be written in terms of the PDFs and partonic 
hard-scattering cross sections as
\begin{align}
\label{eq1}
 \frac{d\sigma}{dM^2}=N{\sum_{ab}}
\int_{\tau}^{1}\frac{dx_1}{x_1} f_a(x_1,\mu)
\int_{\tau/x_1}^{1}\frac{dx_2}{x_2} f_b(x_2,\mu)\,
\omega_{ab}(z,\alpha_s(\mu),r).
\end{align}
The transversely polarized cross section is written in an analogous
manner. In~(\ref{eq1}), $\tau = M^2/S$, $z=\tau/x_1x_2$ and $r=M^2/\mu^2$, 
with $\mu$ the renormalization/factorization scale. 
$N$ is defined so that the ${\cal O}(\alpha_s^0)$
term becomes $\omega^{(0)}_{q\bar{q}}=\delta(1-z)$. The
higher-order functions $\omega^{(i)}_{ab}$ have been calculated
to ${\mathcal O}(\alpha_s^2)$ for the unpolarized
cross section\cite{vN}, and to ${\mathcal O}(\alpha_s^1)$ for the
transversely polarized one\cite{wv}. 

The numerical size of the NNLO corrections for GSI or J-PARC kinematics
amounts to more than three times the LO cross section at high $M$.     
It is known that these large corrections come from the threshold region
where the partonic energy is just enough to produce the lepton pair of
invariant mass $M$. In this region, large ``threshold'' logarithms arise.
The systematic way of taking into account these logarithms to all orders,
called ``threshold resummation'', has been developed in\cite{thre}.
The resummation is achieved in Mellin-moment space, where it
gives rise to a Sudakov exponent. Presently, the exponent for the
Drell-Yan process is known to NNLL accuracy. 
Defining $\omega_{ab}(n)=\int_0^1 dz z^{n-1} \omega_{ab}(z)$, one has:
\begin{align}
 \omega_{q\bar{q}}^{\rm res}(n,\alpha_s,r) &= C_{DY}(\alpha_s,r)
\exp\left[\frac{1}{\alpha_s}h^{(1)}_{q}(\lambda)+h_{q}^{(2)}(\lambda,r)
+\alpha_s h_{q}^{(3)}(\lambda,r)\right],\nn
\end{align}
where $\lambda=b_0\alpha_s\ln{n}$. The detailed expressions for
the $n$-independent coefficient 
$C_{DY}$ and the functions $h^{(i)}_q$ may be found, e.g., 
in Ref.\cite{vogt}. We note that $C_{DY}$ is also known to 
exponentiate\cite{eynck}. We use the ``Minimal
Prescription''\cite{Catani:1996yz} for dealing with the 
Landau pole in the resummed expression.

It is known that the resummation formula can be improved to include
collinear (non-soft) gluon effects\cite{kramer,kulesza}. In NLO, 
these correspond to terms $\propto \ln{(n)}/n$. They may be 
taken into account in the resummation by including certain subleading 
terms in the exponent, associated with DGLAP evolution of parton 
distributions. Through singlet mixing in evolution, these 
subleading terms also feed into the $qg$-subprocess\cite{kulesza}. 
We found that these effects are significant, especially for the case
of $pp$ collisions at J-PARC. 

\begin{figure}[t]
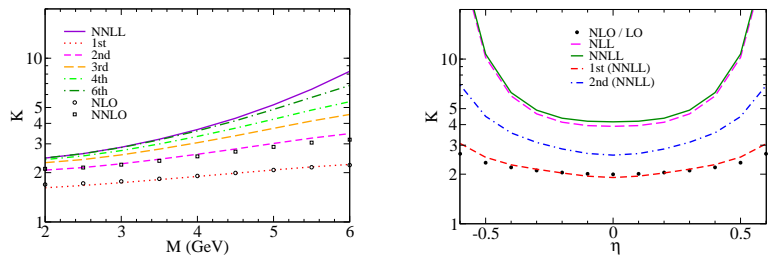

\begin{center}
\includegraphics[width=.4\textwidth]{kfac-mix-exp}
\hspace{20pt}
\includegraphics[width=.4\textwidth]{kfac-cos-5}
\caption{$K$-factors for the resummed cross section and its
perturbative expansions for $pp$ collisions at $\sqrt{S}=10$ GeV. 
Left: invariant-mass distribution, right: rapidity distribution at $M=5$~GeV. 
The NLO (NNLO) $K$-factors are also plotted as circle 
(square) symbols.}\label{kfac}
\end{center}
\end{figure}

Fig.~\ref{kfac}(left) shows the resummed $K$-factor for the J-PARC
situation.  
Expansions of the resummed $K$-factor to fixed perturbative orders
are also plotted. 
We stress that the second- and third-order expansions are in
good agreement with the full NLO and NNLO results. This shows that the
higher-order corrections are indeed dominated by the threshold
logarithms, and that the resummation is accurately reproducing
the latter.   
\section{Resummation for Rapidity Distributions}
We now consider the cross section differential in the lepton
pair's rapidity,  
\begin{align}
\frac{d\sigma}{dM^2d\eta}=N{\sum_{ab}}
\int_{x^0_1}^{1}dx_1 f_a(x_1)
\int_{x^0_2}^{1}dx_2 f_b(x_2)
D_{ab}(x^0_1,x^0_2,x_1,x_2,\alpha_s),
\end{align}
where $x_{1,2}^0=\sqrt{\tau} {\rm e}^{\pm \eta}$. The
$D_{ab}$ have been calculated perturbatively up to ${\cal
O}(\alpha_s^2)$ for unpolarized cross section\cite{anast}, and to ${\cal
O}(\alpha_s)$ for the transversely polarized
case\cite{attnlo}. The ${\cal O}(\alpha_s^0)$
term is simply
$D_{q\bar{q}}^{(0)}=\delta{(x_1-x_1^0)}\delta{(x_2-x_2^0)}$. 
The application of the threshold resummation technique to rapidity
distributions has been discussed e.g.\ in Ref.\cite{sv01}. In addition to 
the usual Mellin transform in $\tau$, it makes use of a Fourier 
transform in $\eta$. The cross section in double-transform space
can be written as  
\begin{align}
\tilde{\sigma}(n,m,\alpha_s,r)&\equiv\int_0^1 d\tau \tau^{n-1}
\int d\eta e^{im\eta}\frac{d\sigma}{dM^2d\eta}\nn\\
&=N{\sum_{ab}}\,
f_a(n+\frac{i}{2}m)f_b(n-\frac{i}{2}m)\tilde{D}_{ab}(n,m,\alpha_s,r).
\end{align}
In the threshold limit, $\tilde{D}_{ab}$ can be
written in terms of the higher-order function $\omega_{ab}(n,\alpha_s)$ 
for the invariant-mass distribution discussed above. The 
resummation may then be performed as before. Details will be presented 
elsewhere. In Fig.~\ref{kfac}(right), we show the $K$-factor for the 
Drell-Yan rapidity
distribution in the J-PARC experiment, at NLO and for the resummed case. 
The $K$-factors increase toward larger $\eta$, since one approaches
the threshold regime more closely there.

\section{Summary}

We have discussed higher-order pQCD effects in the
mass and rapidity distributions for the Drell-Yan process at the 
proposed GSI and J-PARC experiments. The corrections are very large, 
but seem under control when the soft-gluon resummation is implemented.
We hope that our studies, along with the complementary study for 
transverse-momentum distributions\cite{kkst}, will be of use
in comparisons to future data from the GSI and J-PARC.

\section*{Acknowledgments}
We are grateful to H.~Shimizu and G.~Sterman for collaboration
on some of the work described here. H.Y.'s work is supported in 
part by a \uppercase{R}esearch \uppercase{F}ellowship of the 
\uppercase{J}apan \uppercase{S}ociety for
the \uppercase{P}romotion of \uppercase{S}cience.
W.V.\ is supported by DOE Contract No.~DE-AC02-98CH10886.


\begin{thebibliography}{99}
%
\bibitem{gsi} 
V.~Barone {\it et al.}  [PAX Collaboration], arXiv:hep-ex/0505054; \\
M.~Maggiora {\it et al.}  [ASSIA Collaboration], arXiv:hep-ex/0504011.
%
\bibitem{jparc}
D.~Dutta {\it et al.}, Letter of Intent for ``Physics of High-Mass Dimuon
Production at the 50-GeV Proton Synchrotron'';
S.~Sawada, these proceedings.
%
%
%
%
%
%
\bibitem{shimizu}
H.~Shimizu, G.~Sterman, W.~Vogelsang and H.~Yokoya,
Phys.\ Rev.\ D {\bf 71}, 114007 (2005);
Nucl.\ Phys.\ Proc.\ Suppl.\  {\bf 157}, 197 (2006).
%
\bibitem{thre} G.~Sterman, Nucl.\ Phys.\ B {\bf 281}, 310 (1987);
S.~Catani and L.~Trentadue, Nucl.\ Phys.\ B {\bf 327}, 323 (1989); 
{\it ibid.} {\bf 353}, 183 (1991).
%
%
\bibitem{vN} R.~Hamberg {\it et al.}, 
Nucl.\ Phys.\ B {\bf 359}, 343 (1991); ibid.\ B {\bf 644}, 403 (2002). 
%
\bibitem{wv}
W.\ Vogelsang,  Phys.\ Rev.\ D {\bf 57}, 1886 (1998).
%
%
\bibitem{vogt} A.~Vogt,
Phys.\ Lett.\ B {\bf 497}, 228 (2001);
S.~Catani {\it et al.}, JHEP {\bf 0307}, 028 (2003);
S.~Moch and A.~Vogt, Phys.\ Lett.\ B {\bf 631}, 48 (2005).
%
\bibitem{eynck} T.~O.~Eynck, E.~Laenen and L.~Magnea,
JHEP {\bf 0306}, 057 (2003).
%
\bibitem{Catani:1996yz} 
S.~Catani {\it et al.}, Nucl.\ Phys.\ B {\bf 478}, 273 (1996). 
%
\bibitem{kramer}
M.~Kramer, E.~Laenen and M.~Spira, Nucl.\ Phys.\ B {\bf 511}, 523 (1998);
S.~Catani, D.~de Florian and M.~Grazzini, JHEP {\bf 0201}, 015 (2002).
%
\bibitem{kulesza}
  A.~Kulesza, G.~Sterman and W.~Vogelsang,
  Phys.\ Rev.\ D {\bf 66}, 014011 (2002).
%
%
%
%
%
%
\bibitem{anast}
C.~Anastasiou {\it et al.}, Phys.\ Rev.\ D {\bf 69}, 094008 (2004).
%
\bibitem{attnlo} 
O.~Martin {\it et al.}, 
Phys.\ Rev.\ D {\bf 60}, 117502 (1999).
%
%
\bibitem{sv01}
G.~Sterman and W.~Vogelsang, JHEP {\bf 0102}, 016 (2001);\\
A.~Mukherjee and W.~Vogelsang, Phys.\ Rev.\ D {\bf 73}, 074005 (2006).
%
\bibitem{kkst}
H.~Kawamura {\it et al.}, 
Prog.\ Theor.\ Phys.\  {\bf 115}, 667 (2006); these proceedings.
%
\end{thebibliography}
\end{document}